\documentclass[twocolumn,letterpaper,aps,prb,floats,floatfix]{revtex4}
\usepackage{amsmath}
\usepackage{amssymb}
\usepackage{amsfonts}
\usepackage{psfig}
\usepackage{epsfig}
\usepackage{graphicx}
\usepackage{dcolumn}

\def\ham{\mathcal{H}}
\def\iGLE{\lowercase{i}GLE\ }
\begin{document}
\preprint{\scriptsize JOURNAL OF CHEMICAL PHYSICS;\hspace{4pt}
to be submitted October, 2004}
%
\title{The projection of a nonlocal mechanical system 
onto the irreversible generalized Langevin equation, II: 
Numerical simulations}
\author{Marc Vogt}
\thanks{Present address: Department of Biochemistry and Molecular Biology,
University of Massachusetts, 913 Lederle Graduate Research Tower,
710 North Pleasant Street, Amherst, MA 01003}
\author{Rigoberto Hernandez}
\thanks{Author to whom correspondence should be addressed}
\email{hernandez@chemistry.gatech.edu.}
\affiliation{%
Center for Computational and Molecular Science and Technology, \\
School of Chemistry and Biochemistry\\
Georgia Institute of Technology\\
Atlanta, GA 30332-0400}
\date{\today}
\begin{abstract}%
The irreversible generalized Langevin equation (iGLE)
contains a nonstationary friction kernel that 
in certain limits reduces to the GLE with space-dependent friction.
For more general forms of the friction kernel, the iGLE was
previously shown to be the projection of a mechanical system
with a time-dependent Hamiltonian.
[R.~Hernandez, {\it J.~Chem.~Phys.~\bf 110}, 7701 (1999)]
In the present work, 
the corresponding open Hamiltonian system is further explored.
Numerical simulations of this mechanical system 
illustrate that the time dependence of the
observed total energy and the correlations of the solvent force
are in precise agreement with the projected iGLE.
\end{abstract}
\maketitle


\section{Introduction}


Several models for stochastic motion involve the Langevin
and generalized Langevin equation (GLE) in so far as the environment
is assumed to be in a steady-state equilibrium throughout the
dynamical event, {\it vis-a-vis} the linear responding bath
is stationary and obeys a fluctuation-dissipation 
theorem.\cite{kram40,zwan60b,zwan61b,prig61,ford65,mori65,kubo66,hynes85a,hynes85b,nitzan88,berne88,rmp90,tucker91,tucker93,pollak96}
In recent work,\cite{hern99a,hern99b,hern99c,hern99d,hern99e}
it was suggested that in many instances, the environment is
not stationary, and as such a nonstationary version of the
GLE would be desirable.
In principle, it is easy to write a nonstationary GLE as
\begin{subequations}\label{eq:nGLE}
\begin{equation}
{\dot v} =
 - \int^tdt' \gamma(t,t'){v}(t')
 +\xi(t) + F(t),\label{gle}
\end{equation}
where $F(t)$ ($\equiv -\nabla_qV(q(t))$) is the external force,
$v$ ($=\dot q$) is the velocity, $q$ is the mass-weighted
position, $\xi(t)$ is the random force
due to a solvent, and $\gamma(t,t')$ 
represents the nonstationary response of the solvent.
To completely specify this system of equations, a closure connecting
the random force to the friction kernel is needed.
The fluctuation-dissipation relation (FDR) provides such a
closure for the GLE.\cite{kubo66}
An obvious generalization of the FDR for nonstationary
friction kernels is the requirement,
\begin{equation}
\gamma(t,t') = \langle \xi(t) \xi(t') \rangle\;.
\label{eq:ngle_fdr}
\end{equation}\end{subequations}%
Unfortunately, such a construction will not necessarily be
satisfied for an arbitrary nonstationary friction kernel,
nor will it necessarily be associated with the dynamics of
a larger mechanical system.

The GLE model with space-dependent friction (xGLE)
developed by Lindenberg and coworkers,\cite{lind81,lind84} 
and Carmeli and Nitzan\cite{carm83} 
does exhibit the structure of Eq.~(\ref{eq:nGLE})
and as such is an example of nonstationary stochastic dynamics.
In recent work,\cite{hern99a,hern99b,hern99c,hern99d,hern99e}
a generalization of this model was developed 
which includes arbitrary nonstationary changes in the strength
of the friction, but like the xGLE model does not include
changes in the response time. 
As such it is not quite the ultimately desired nonstationary GLE.
Avoiding redundancy in the term ``generalized GLE,'' the
new formalism has been labeled the {\it irreversible} 
generalized Langevin equation (iGLE), thereby emphasizing the
irreversibility---vis-a-vis nonstationarity---in 
the response of the quasi-equilibrium environment.
But this ``irreversibility" may not persist at long times, and therefore
Drozdov and Tucker chose to call such an equation of motion
the multiple time-scale generalized Langevin equation (MTSGLE) in
their application of the iGLE to study local density enhancements 
in supercritical fluids.\cite{tucker01}

In this paper, the iGLE model is first summarized in Sec.~\ref{sec:iGLE},
explicitly indicating the limit in which position-dependent friction may be
recovered.
In earlier work,\cite{hern99e} (Paper I) it was shown that the iGLE may
be obtained as a projection of an open Hamiltonian system,
in analogy with the similar construction for the 
GLE.\cite{ford65,zwan73,caldeira81,cort85,pollak86b}
In Sec.~\ref{sec:iGLE}, the connections between the projection of 
the Hamiltonian of Paper I onto a chosen dynamical variable
and that obtained by the xGLE are further illustrated,
and the possibly troubling nonlocal term it contains is
also further explored.
The results of several numerical simulations of the Hamiltonian system
are presented in Sec.~\ref{sec:sim} in order to illustrate
the effect of the nonlocal term on the dynamics, and to 
verify that the constructed random force does obey the
FDR.

\section{The \iGLE and Space-Dependent Friction}\label{sec:iGLE}

\subsection{Stochastic Dynamics}

The iGLE may be written as 
\begin{equation}
{\dot v}(t) =
 - \int_0^t\!dt'\, g(t) g(t') \gamma_0(t-t') {v}(t')
 +g(t) \xi_0(t)  + F(t)\;,
\label{eq:ggle}
\end{equation}
where $g(t)$ characterizes the irreversibility in the
equilibrium environment, and there exists a FDR
between the Gaussian random force $\xi_0(t)$ and
the stationary friction kernel $\gamma_0(t-t')$.
Through the identities,
\begin{subequations}\begin{eqnarray}
\gamma(t,t') &\equiv& g(t) g(t') \gamma_0(t-t')\\
\xi(t) &\equiv& \xi_0(t)\;,
\end{eqnarray}\end{subequations}%
the iGLE is a construction of the nonstationary Eq.~(\ref{eq:nGLE}).
One possible interpretation of the role of $g(t)$ in the
iGLE is that it corresponds to the strength of the 
environmental response as the reactive system traverses
the environment through an {\it a priori} 
specified trajectory, call it $y(t)$.  
Assuming that one also knows the field $f(y)$, which
is the strength of the environmental friction over this configuration
space, then the irreversibility may be written as,
\begin{equation}
g(t) = f(y(t))\;.\label{eq:xGLE}
\end{equation}
In the case that the chosen coordinate is itself the configuration
space over which the friction varies ---{\it i.e.}, $y=x$---
the GLE with space dependent friction of Carmeli and Nitzan is 
formally recovered.\cite{carm83}

But the iGLE is more general than the xGLE
because it allows for a variety of ``irreversible" time-dependent
environments.
For example, in the WiGLE model,\cite{hern99d} 
each particle ---labeled by $n$---
is in an environment induced by the average of
the properties of itself and $w$ neighbors,
\begin{subequations}\begin{eqnarray}\label{eq:wigle-ga}
g_n(t) & \equiv & \left< \left| R(t) \right| \right>_n^\zeta \\
\left< \left| R(t) \right| \right>_n & \equiv & \frac{1}{w+1} 
\sum_{i\in S_{w,n}} \left| R_i(t) \right| \;, \label{eq:wigle-gb}
\end{eqnarray}\end{subequations}%
where $S_{w,n}$ is the set of $w$ neighbors.
In the limit that $w\to0$, the chosen coordinate is dissipated
only by a function of its position, which is precisely the
limit of space-dependent friction.
In the limit that $w\to\infty$, the chosen coordinate is
instead dissipated by a macroscopic average of the motion of all
the reacting systems in the sample.
The contribution of a particular particle to this average is
infinitesimally small, and hence the
friction contains no space-dependent friction.
In between these limits, there is a balance between self-dissipation
due to a space-dependent friction term, and heterogeneous
dissipation due to the average of the motion of the 
$w$ neighbors.

\subsection{Mechanical Systems}

In recent work, a Hamiltonian has been obtained whose projection
is the iGLE when $g(t)$ depends exclusively on time
---{\it i.e.}, it includes neither explicit space-dependence
nor the WiGLE dependence.\cite{hern99e}
This so-called iGLE Hamiltonian may be written as
\begin{subequations}\label{eq:iGLEall}\begin{eqnarray}
\ham_{\rm iGLE}
&=&    \frac{1}{2}p_q^2 + 
     \left\{ V(q) + \delta V_1(q,t) + \delta V_2[q(\cdot),t] \right\}
\nonumber\\ &&\quad   
 - g(t) \left[\sum_{i=1}^Nc_i x_i\right]q
\nonumber\\ &&\quad   
 + \sum_{i=1}^N\left[\tfrac{1}{2}p_i^2 + \tfrac{1}{2} \omega_i^2x_i^2\right]\;,
\label{eq:iGLEham}
\end{eqnarray}
where 
\begin{eqnarray}
\delta V_1(q,t) &\equiv& 
  \tfrac{1}{2} g(t)^2\sum_{i=1}^N 
    \frac{c_i^2 }{ \omega_i^{2}} q^2 
\label{eq:Vzero}\\
\delta V_2[q(\cdot),t] &\equiv& 
\tfrac{1}{2} \int_0^t\!dt'\,a(t,t')\left[q(t')-q(t)\right]^2
\nonumber\\ &&\quad   
- \tfrac{1}{2} \left[\int_0^t\!dt'\,a(t,t')\right] q(t)^2
\;,
\label{eq:Vone}
\end{eqnarray}\end{subequations}%
where
\begin{equation}
a(t,t')\equiv g(t)\dot g(t')\gamma_0(t-t')\;,
\end{equation}
and the time dependence in $q(t)$ is explicitly included in the
definition of the $\delta V_2[q(\cdot),t]$ functional for clarity.
Ignoring the $\delta V_2$ term and identifying $g$ as in Eq.~(\ref{eq:xGLE}),
this Hamiltonian is similar to the xGLE Hamiltonian for
space-dependent friction.
This result is not surprising in the sense that the iGLE has a similar 
generalized structure.
However, the xGLE Hamiltonian gives rise to an additional dependence
on $q$ whereas the iGLE Hamiltonian gives rise to an additional
dependence on time $t$.  The projections are thus analogous but
not exactly the same.

\subsection{Equation of Motion}

The nonlocality in the $\delta V_2[q(\cdot),t]$
term does present some difficulties
which are worth considering.
In the absence of this term, the extremization of the action
readily leads to the usual Hamilton's equations.
In general, the presence of the $\delta V_2$ term
contributes to the time evolution of the momentum,
$\dot p$, by what of the functional derivative,
\begin{equation}
-\frac{\delta S_2 }{ \delta q(t)}\;,
\end{equation}
where the contribution to the action due to $\delta V_2$
may be written as
\begin{eqnarray}
S_2 &\equiv&
\frac{1}{2} \int_0^T\!dt\,
      \int_0^t\!dt'\,a(t,t')q(t')^2 
\nonumber\\&&
- \int_0^T\!dt\,
      \int_0^t\!dt'\,a(t,t')q(t)q(t')\;,
\end{eqnarray}
and $T$ is the arbitrary final time to which the action is evaluated.
A simple calculation readily leads to
\begin{eqnarray}
-\frac{\delta S_2 }{ \delta q(t)} &=&
  \int_0^t\!dt'\,a(t,t')q(t')
\nonumber\\&&
+ \int_t^T\!dt'\,a(t,t')\left[q(t')-q(t)\right]\;.
\end{eqnarray}
The first time in the RHS precisely cancels the contribution
due to the nonstationarity in the friction kernel.
However, the remaining second term depends on the arbitrary
final time $T$.
It's presence can't be physically correct
because it leads to different dynamics depending
on the choice of $T$.
In the limit that $T$ is near ---though greater--- than $t$, this term
vanishes, however.  
This suggests that an additional approximation 
ignoring the second term, thereby eliminating the transient
effects from a term that depends arbitrarily on $T$, is warranted.
(And this is consistent with the Carmeli and Nitzan derivation, in
that they too need to remove transient terms.)
Within this approximation, the projection in Ref.~\onlinecite{hern99e},
then leads to the iGLE.

Thus the projection of the iGLE Hamiltonian 
leading to the iGLE 
with a purely time-dependent friction 
is analogous (\& complementary) to the Carmeli \& Nitzan projection 
to a GLE with space-dependent friction.
The construction of such a Hamiltonian for an iGLE with arbitrary
nonstationary friction, as manifested in WiGLE, is still an open
problem.
In the next section, the dynamics of the iGLE
Hamiltonian is explored
through numerical simulations in order to observe
the degree of energy conservation ---as it is affected by
$\delta V_2$--- and the correlation of the 
constructed forces.


\begin{figure}
\includegraphics*[width=2.0in,angle=-90]{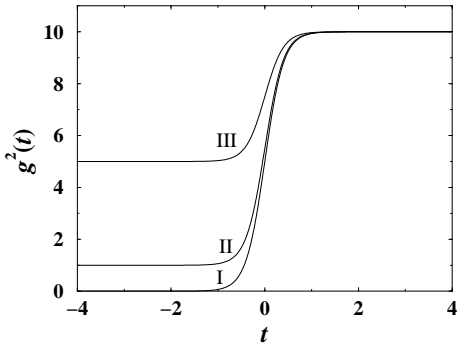}
\caption[Switching function for the system friction.]%
{The square of the irreversible change in the environment
$g(t)^2$ is shown here as a function of time for the three cases
examined in this study.}
\label{fig:switch}
\end{figure}
\begin{figure}
\includegraphics*[width=2.5in]{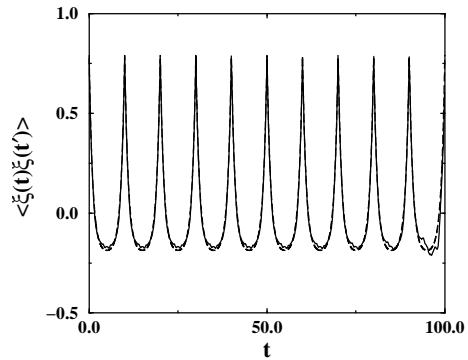}
\caption
{The average correlation of the random forces on the tagged particle $\langle \xi (t) \xi(t^{\prime}) \rangle$
as a function of time $t$ (solid line) for the GLE case with coupling
constants calculated as in Eqn.{tuckercrazy}.  The dashed line represents
the friction kernel as calculated in Eqn.{eq:freq}.
}
\label{fig:gleforces}
\end{figure}

\section{Numerical Results}\label{sec:sim}
 
In this section, numerical simulations of the Hamiltonian
equivalent of the iGLE are presented.  It is shown that the 
inclusion of the non-local term,
$\delta V_2[q(\cdot),t]$, known to be small in the quasi-equilibrium
regime is actually necessary generally.  That is, 
throughout this work, the value of
$\delta V_2[q(\cdot),t]$ is non-zero during the
increase in system coupling.
The latter is  specified through the function
$g(t)$ chosen to satisfy
\begin{subequations}\label{eq:switch}
\begin{eqnarray}
g(t)^2 &=& g(-\infty)^2 + \tfrac{1}{2}[g(\infty)^2 - g(-\infty)^2]
\nonumber\\ &&\quad   
\times \biggl( 1 + \frac{e^{t/{\tau_{g}} -1}}{e^{t/{\tau_{g}} +1}} \biggr),
\end{eqnarray}
as illustrated in Fig.~\ref{fig:switch},
or equivalently as
\begin{eqnarray}
g(t)^2 &=& \tfrac{1}{2}[g(\infty)^2 + g(-\infty)^2] 
\nonumber\\ 
&&+ \tfrac{1}{2}[g(\infty)^2 - g(-\infty)^2]\tanh(\tau/\tau_{g}).
\end{eqnarray}
\end{subequations}


\subsection{Coupling Constants}

The values of the coupling constants 
in Eqn.~\ref{eq:iGLEham} can be obtained from
the reverse cosine transform of Eqn.~\ref{eq:freq},
\begin{eqnarray}
\frac{c_{i}^2}{\omega_{i}^2} = 
  \frac{2 \omega_c}{\pi} 
  \int_{0}^{\infty}dt \cos(\omega_i t)\gamma_{0} e^{-t/\tau},
\end{eqnarray}
such that an effective (discretized) friction kernel of the form,
\begin{equation}
\gamma_{0}(t) = \sum_{i=1}^{N} \frac{c_{i}^2}{\omega_{i}^2} \cos(\omega_{i}t).
\label{eq:freq}
\end{equation}
is used to approximate $\gamma_{0} e^{-t/\tau}$.
The coupling constants $c_i$ in Eqn.~\ref{eq:iGLEham} 
are readily obtained on a discrete (and finite) grid of
$N$ nontrivial frequencies, $\omega_i\equiv i\omega_c$, by
equating the spectral densities of the exponential friction to 
that of discretization.
The smallest frequency $\omega_c\equiv1/(M\tau)$ can be characterized
in terms of the charateristic integer $M$ which effectively sets 
the longest time scale that can be measured before false recurrences
appear due to the discretization.
The coefficients can be written (as, for example, also obtained 
by Topaler and Makri\cite{makri94}) as
\begin{eqnarray}
c_{i}=\sqrt{(\frac{2 \gamma_{0} \tau \omega_{c}}{\pi})\cdot 
(\frac{\omega_{i}^2}{1+\omega_{i}^2 \tau^2})
}\;.
\label{eq:coupl}
\end{eqnarray}
$1/(N\omega_{c})$ represents the shortest time scale of interest.
This connection between the continuum stationary friction
kernel and the discrete number of frequencies and coupling strengths 
(Eq.~\ref{eq:freq})
is exact in the continuum limit ($N \rightarrow \infty$).  
However, $M$ must also be large enough so as to 
ensure the decay in correlations between the
bath modes; typically $M\ge4$.
Simulations of the GLE (with constant friction) were performed to confirm
a suitable number of bath particles for which convergence of
the relationship in Eqn.~\ref{eq:freq} was observed.  
It was found that as few as 20 harmonic bath modes 
can yield acceptable convergence of the
velocity correlation function of the chosen coordinate
because its decay is much faster than the recurrence time in the 
bath modes.
Nonetheless, to ensure that there are enough modes
to approximately satisfy the continuum limit at the longest 
and shortest times of interest while also limiting the 
requisite computing power, the number of harmonic bath modes, $N$,
used in the present work have been taken to be 200.


Although the coupling constants have been calculated 
as per Eq.~\ref{eq:coupl} in the simulations
of the mechanical projection of the iGLE presented in this work,
it is beneficial to examine some other choices of the 
coupling constants, $c_i$.
The main question is how to best equate the continuum (left hand side)
and discrete (right hand side) representations of Eqn.~\ref{eq:freq} in
the frequency domain.
One alternative method is that of Tucker and coworkers,\cite{reese2}
in which
the coupling constants are obtained, not by integrating the spectral function
over an infinite domain as above, but over a domain bounded by the
longest time scale $1/{\omega_c^{'}}$.  
The resulting coefficients,
\begin{eqnarray}
c_i = \sqrt{(\tfrac{2}{t_c} \omega_{i}^{2} \gamma_{0})
   \Biggl[
     \frac{1/{\tau}}{\omega_{i}^{2} + 1/{{\tau}^{2}}} + 
     \frac{e^{-t_{c}/{\tau}}\omega_{i}\sin{(\omega_{i}t_{c})}}%
          {\omega_{i}^{2}+ 1/{{\tau}^{2}}}  
   \Biggr]},
\label{tuckercrazy}
\end{eqnarray}
are associated with a correspdonging discrete frequency
$\omega_i\equiv i\omega_c^{'}$ as before, but the smallest
frequency is redefined as $\omega_{c}^{'} = \pi /(p \tau)$
for some characteristic integer $p$.
The use of the coupling constants of Eq.~\ref{tuckercrazy}
in the Hamiltonian representation of the GLE
yields a very good match between the friction kernel
as specified by Eqn.~\ref{eq:freq} and the correlation function for the random
forces on the tagged particle as shown in Fig.~\ref{fig:gleforces}.
Unfortunately, the system still retains a long-time periodicity associated
with the mode described by the largest inverse frequency.
%
%
In an attempt to sidestep this issue, the chosen frequencies in the domain
could be chosen incommensurably by choosing a frequency randomly within each
window, $I_i\equiv ( \{i-1\}\omega_c, i\omega_c]$.
Given a random sequence of frequencies $\{\omega_i^{\rm R}\}$ 
in which $\omega_i^{\rm R}\in I_i$ for each $i$,
the coefficients can then be re-evaluated leading to the result,
\begin{eqnarray}
c_{i} &=& \Biggl\{(\tfrac{2 \gamma_{0}}{\pi}) 
            \left( {\omega_{i}^R}
           \right)^{2} 
   \Biggl[\tanh^{-1}( i\tau\omega_c )
\Biggr.\Biggr.\nonumber\\ \Biggl.\Biggl.
    &&\quad  +   \tanh^{-1}( \{i-1\}\tau\omega_c)
   \Biggr]\Biggr\}^{\frac{1}{2}}
\;.
\end{eqnarray}
This choice of coefficients was also tested on the GLE but it 
led to similar results for the 
correlation function of the projected (random) forces $\xi(t)$,
both in terms of the accuracy
and in reproducing the long-time periodicity.  
It was therefore determined that the results for the GLE
(and thereby the iGLE)
are not highly sensitive to the limiting form of the coupling constants 
so long as the choice satisfies the appropriate friction kernel,
while the random choice of frequencies did not remove the false long-time
periodicity in the autocorrelation of the force $\xi(t)$.

This discussion, though somewhat pedantic, does offer a critical warning:
any numerically measured behavior in the chosen particle that is
correlated for times longer than the period of the false long-time
periodicity in the discrete representation is suspect to error.
This, in turn, places an upper bound on the slowest time scale 
---{\it viz.}~$t_g$ in Eq.~\ref{eq:switch}--- that
can be imposed on the nonstationary behavior of the bath coupling
for a given choice of discrete oscillators in the Hamiltonian representation.
In the simulations that follow, this bound is, indeed satisfied.

\subsection{The Free Particle}

The numerical equivalence between the
iGLE and the Hamiltonian system of Eq.~\ref{eq:iGLEall}
can be illustrated using the 
same model of the nonequilibirium change in the environment
originally investigated in the context of the 
phenomenological iGLE.\cite{hern99a}
To this end, numerical results are presented for the Hamiltonian
system of a tagged free particle ($V(q) = 0$) bilinearly coupled to
a bath of 200 harmonic modes whose smallest
characteristic bath frequency is $\omega_c = 1/(M \tau)$
where $M = 4$.  
Individual bath frequencies are taken at discrete values,
$\omega_{i} = (i- 1/2) \omega_{c}$, and coefficients as per 
Eq.~\ref{eq:coupl}.
All other parameters have identical values to those used in the
numerical integration studies of the iGLE in Ref.~\onlinecite{hern99a}
with the exception that 100,000
trajectories were used in this study, which is a 10-fold increase.
In summary, all simulations share the following
set of parameters: $N = 100,000, k_{B}T = 2.0, \gamma_{0} = 10.0,
\tau = 0.5, \tau_{g} = 0.2, \Delta t = 1$x$10^{-4} (t \geq -8)$
and $\Delta t = 1$x$10^{-3} (t < -8).$
The time dependent friction is modulated 
through the switching function $g(t)$ as shown in Fig.~\ref{fig:switch}.
Each individual trajectory is first equilibrated from $t=-20$ to
ensure that the observed dynamics are influenced only by the
irreversible change in the friction and not the dynamics of 
equilibration.  The system is propagated using the velocity-Verlet
method with smaller timesteps during the regime of friction increase
than for the constant friction regimes.\cite{allen87}

\begin{figure}
\includegraphics*[width=2.5in]{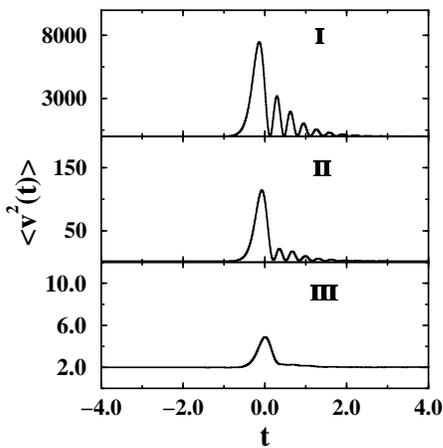}
\caption
{The mean square velocity $\langle v^2(t) \rangle$ is
displayed for each of the cases as a function of time t
and case I, II or III with the non-local term
$\delta V_2[q(\cdot),t]$ set equal to zero.
}
\label{fig:vsqnoV2}
\end{figure}

In order to determine whether or not the 
$\delta V_2[q(\cdot),t]$ term is negligible,
simulations were first performed with $\delta V_2[q(\cdot),t] = 0$.
For all three cases of the change in friction, the non-local
$\delta V_2[q(\cdot),t]$ term was found to be non-negligible
as can be seen in  Fig.~\ref{fig:vsqnoV2} by the fact that 
the system does not obey equipartition of energy during times 
near $t=0$.  
The larger the friction increase, the larger the deviation
from equipartition.  This can be seen in Fig.~\ref{fig:vsqnoV2}
as the largest deviations are seen for case I where the switching function
$g(t)$ increases from 0 to 10.  The friction increases along a similar range
and the average square velocity $\langle v(t)^2 \rangle$ peaks
at near 8000 around $t=0$.  The system does not conserve
total energy in these cases.

\begin{figure}
\includegraphics*[width=2.5in]{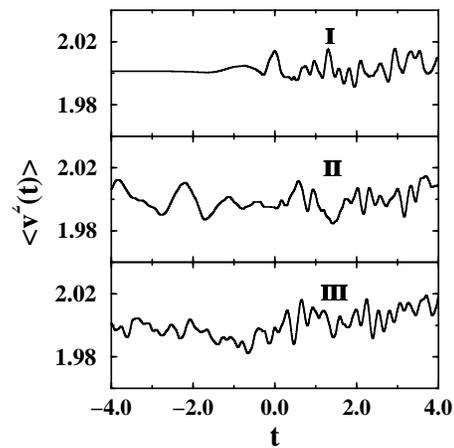}
\caption
{The mean square velocity $\langle v^2(t) \rangle$ is
displayed for each of the cases as a function of time t
and case I, II or III with the non-local term
$\delta V_2[q(\cdot),t]$ explicitly considered.
}
\label{fig:vsq}
\end{figure}

However, upon introducing the nontrivial terms,
$\delta V_2[q(\cdot),t]$ and its derivative (Eq.~\ref{eq:Vone}),  
within the Hamiltonian equations of motion,
equipartition for the system is preserved, as shown in Fig.~\ref{fig:vsq}.  
The $\delta V_2[q(\cdot),t]$ term is therefore not negligible in the extreme
test cases examined in this work in which all the interesting
time scales $1/\gamma_0$, $t_g$, and $\tau$ are 
comparable.
Explicit integration
of $\delta V_2[q(\cdot),t]$ is computationally expensive.
This expense can be reduced when the 
stationary friction kernel has an exponetial form.
In this case, the derivative,
$\frac{\partial{\delta V_2[q(\cdot),t]}}{\partial t}$,
can be calculated using the auxiliary variable
$z$, where
\begin{eqnarray}
\frac{\partial z}{\partial t} = \dot{g}(t) q(t) - \tfrac{1}{\tau} z(t).
\end{eqnarray}
The value of $\frac{\partial{\delta V_2[q(\cdot),t]}}{\partial t}$ 
can then be obtained at each timestep,
\begin{eqnarray}
\frac{\partial{\delta V_2[q(\cdot),t]}}{\partial t} = \gamma_{0}g(t)z(t).
\end{eqnarray}
%
All of the results shown in this paper have been calculated using the 
integration of the auxiliary variable $z$ because it is 
consequently formally equivalent to the explicit integration of Eq.~\ref{eq:Vone} 
while requiring fewer computing cycles.  

As can be seen from Fig.~\ref{fig:vsq},
all three test cases lead to the same level of fluctuations in the long time regime
even though they begin from different initial states.  In the limit
of an infinite number of
trajectories, all cases would exhibit no fluctuations, but these
fluctuations due to finite size effects are indicative of the system
dynamics.  For example, in case I the particle obeys equipartition
perfectly for the early regime, because its motion is ballistic,
whereas the early time dynamics for cases II and III clearly show the influence 
of coupling to the harmonic bath particles.  It can be seen 
from the plot of the average square velocity that 
the system conserves energy in these calculations
when the $\delta V_2[q(\cdot),t]$ term is incluced.
Unfortunately the addttional test for the conservation of energy
cannot be computed directly in these cases because
$\delta V_2[q(\cdot),t]$ depends on the function q(t)
which cannot be known {\it a priori}.

\begin{figure}
\includegraphics*[width=2.5in]{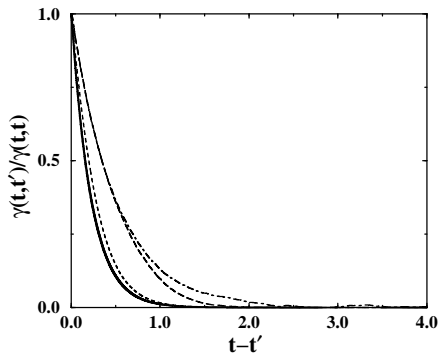}
\caption
{The nonstationary friction kernel $\gamma (t,t^{\prime})$
in case I at several different times $t$
(-4.0 thick solid line, -0.5 solid line, 0.0 dashed line,
1.0 long dashed line and 4.0 dot dashed line)
as a function of the
previous times $t-t^{\prime}$.  The friction kernel is normalized
by the value $\gamma (t,t)$ for illustrative
and comparative  purposes.
}
\label{fig:fric1}
\end{figure}

\begin{figure}
\includegraphics*[width=2.5in]{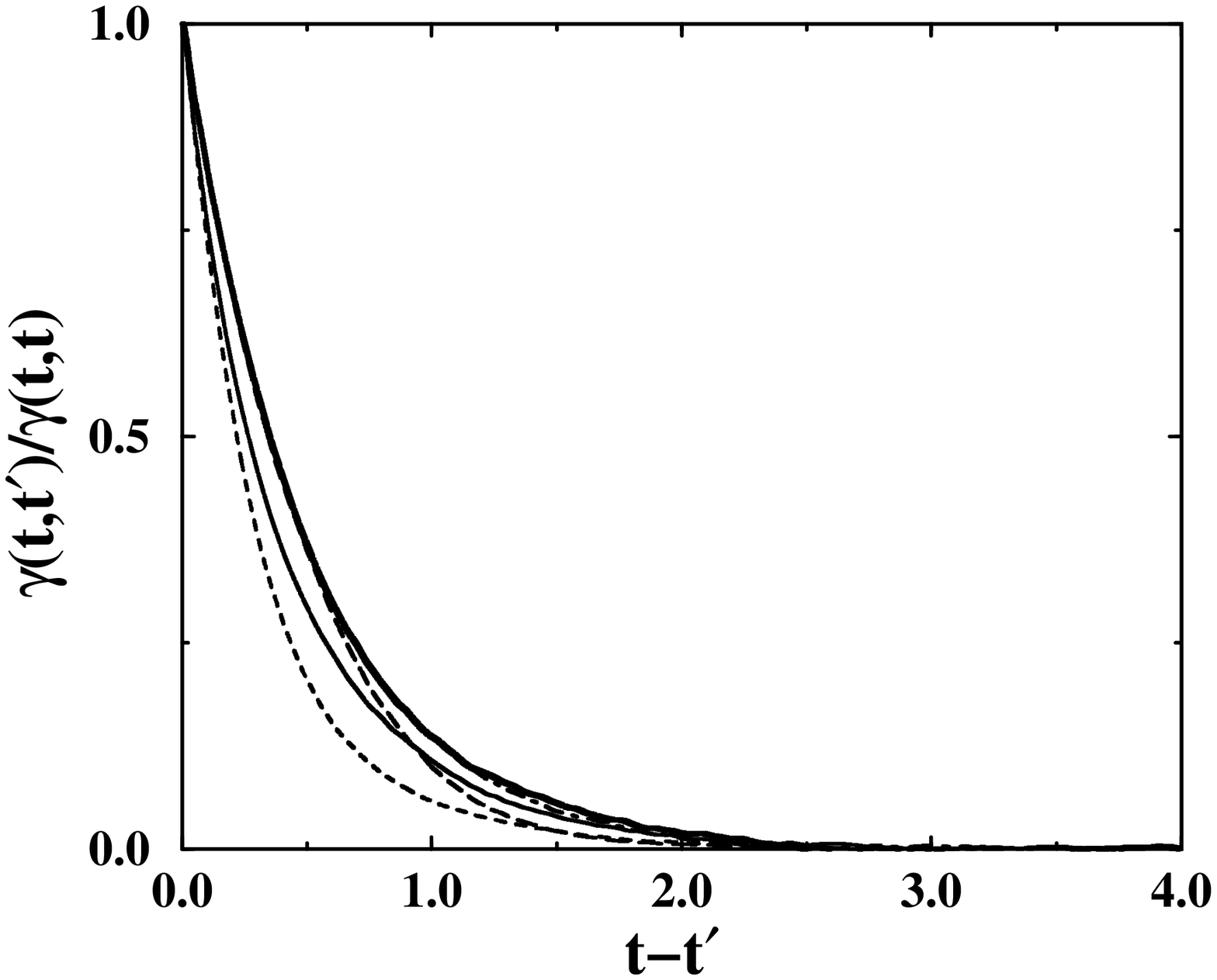}
\caption
{The nonstationary friction kernel $\gamma (t,t^{\prime})$
in case II at several different times $t$
(-4.0 thick solid line, -0.5 solid line, 0.0 dashed line,
1.0 long dashed line and 4.0 dot dashed line)
as a function of the
previous times $t-t^{\prime}$.  The friction kernel is normalized
by the value $\gamma (t,t)$ for illustrative
and comparative  purposes.
}
\label{fig:fric2}
\end{figure}

\begin{figure}
\includegraphics*[width=2.5in]{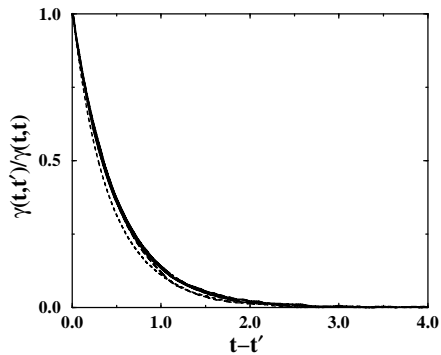}
\caption
{The nonstationary friction kernel $\gamma (t,t^{\prime})$
in case III at several different times $t$
(-4.0 thick solid line, -0.5 solid line, 0.0 dashed line,
1.0 long dashed line and 4.0 dot dashed line)
as a function of the
previous times $t-t^{\prime}$.  The friction kernel is normalized
by the value $\gamma (t,t)$ for illustrative
and comparative  purposes.
}
\label{fig:fric3}
\end{figure}

The construction of the iGLE also requires that the correlation
function of the random forces satisfies a nonstationary extension
of the FDR,
\begin{eqnarray}
k_{B}Tg(t)g(t^{\prime})\gamma_{0}(t-t') = \langle \xi(t) \xi(t') \rangle,
\end{eqnarray}
with respect to the explicit forces $\xi(t)$ on the tagged particle
seen in a microscopic system,
\begin{eqnarray}
\xi(t)= \dot{v} + \gamma_{0}g(t)\int_0^t\!dt'\, g(t') \gamma_0(t-t') {v}(t') - F(t).
\end{eqnarray}
That the mechanical system satisfies this relationship is
illustrated in
Fig.~\ref{fig:fric1}, Fig.~\ref{fig:fric2} and Fig.~\ref{fig:fric3}
for cases I, II and II respectively.
The integral can be simplified 
using an auxiliary variable $z_{2}$ akin to the method for replacing
${\delta V_2[q(\cdot),t]}$ with $z$. 
The auxiliary variable satisfies,
\begin{eqnarray}
\frac{\partial z_{2}}{\partial t} = \dot{g}(t) v(t) - \tfrac{1}{\tau} z_{2}(t).
\end{eqnarray}
The explicit expression for 
the forces on the tagged particle can then be obtained
at each timestep by substitution,
\begin{eqnarray}
\xi(t)= \dot{v} + \gamma_{0}g(t)z_{2}(t) - F(t).
\end{eqnarray}
The autocorrelation function of the force due to the
bath ---{\it viz.}, the random force in the projected variables---
obey the FDR for all cases and times.
It should be emphasized that this remarkable agreement would not have been
found if the nonlocal term were omitted.

\begin{figure}
\includegraphics*[width=2.5in]{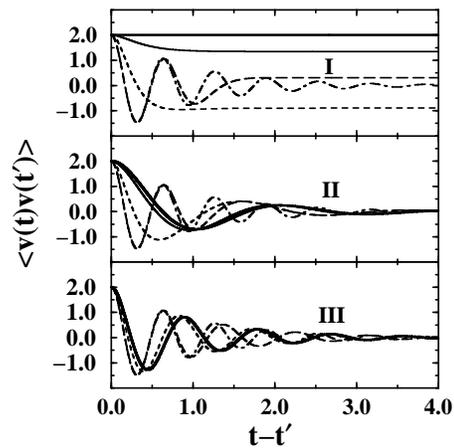}
\caption
{The autocorrelation function of the velocity
of the free tagged particle, $\langle v(t)v(t^{\prime}) \rangle$,
is displayed for each of the three cases of the change in friction
at initial times $t$=
(-4.0 thick solid line, -0.5 solid line, 0.0 dashed line,
1.0 long dashed line and 4.0 dot dashed line).  The panels
indicate cases I,II and III from top to bottom.
}
\label{fig:vauto}
\end{figure}

The velocity autocorrelation functions for the tagged free
particle are shown in 
Fig.~\ref{fig:vauto}.  
The dynamics of the chosen coordinate changes in the
same manner as for the studies on the numerical integration of the iGLE.  
The curves match the previous results exactly with the exception
that the $t=0$ and $t=-0.5$ were mislabeled in Ref. \onlinecite{hern99a}.
The long time ($t= 4.0$) autocorrelation functions
are all the same indicating that all cases reach the same equilibrium
as would be expected.  In case I, the early time (t= -4.0)
autocorrelation function is a straight line since the particle 
is in the ballistic regime.  All curves start at approximately 2.0
for $t-t^{\prime} = 0$ since the system satisfies equipartition.

\subsection{The Particle in a Harmonic Potential}

\begin{figure}
\includegraphics*[width=2.5in]{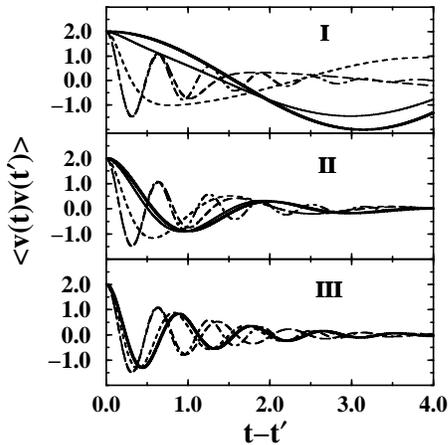}
\caption
{The autocorrelation function of the velocity
of the harmonic tagged particle, $\langle v(t)v(t^{\prime}) \rangle$,
is displayed for each of the three cases of the change in friction
at initial times $t$=
(-4.0 thick solid line, -0.5 solid line, 0.0 dashed line,
1.0 long dashed line and 4.0 dot dashed line).  The panels
indicate cases I,II and III from top to bottom.
}
\label{fig:vautoh}
\end{figure}

The same simulations were run for a tagged particle in a harmonic well
characterized by a frequency $\omega=1$.
The results are essentially all the same with the following exceptions:
The system with tagged particle in a harmonic well is less
sensitive to the increase in friction and can be simulated accurately
with larger timesteps than the free particle case due to the confining
effect of the harmonic well.  Similarly, the system does not yield 
such large spikes in the average square velocity for simulations 
with $\delta V_2[q(\cdot),t] = 0$, although the explicit calculation of that
term is still a necessity for accurate results.  
Clearly the early time dynamics will differ between the free particle
and harmonic particle cases as can be seen from the velocity autocorrelation
functions for the harmonic tagged particle case in
Fig.~\ref{fig:vautoh}.
Interestingly though, this is only the case
for case I.  In cases II and III the dynamics are essentially 
identical to the free particle case.  This is due to the fact that
the initial friction and thus the coupling is so strong
(10.0 for case I and 50.0 for case II) that the tagged particle
potential has an insignificant contribution to the
energy as compared to the contribution from the interaction
with the bath.

\section{Concluding Remarks}

The equivalence of the stochastic
iGLE and the deterministic Hamiltonian system has been
demonstrated by numerical integration of the equations of motion
corresponding to the Hamiltonian in Eqn.~\ref{eq:iGLEham}.
The Hamiltonian system with 200 bath particles
has been shown to exhibit equivalent dynamics as that seen in
numerical integration of the iGLE.  This is expected to be the
case in the infinitely large bath size regime.  The equivalence
is contingent upon the explicit evaluation of the non-local
dissipative term which is non-zero for all the cases
of interest in this study.  Although the free particle and harmonic
oscillator cases do not contain a reactant/product boundary,
they should be sufficient to verify the general agreement between the iGLE
and the mechanical oscillator system.
The deterministic mechanical system
serves as further
evidence that the stochastic equation of motion is not purely a
fiction (e.g., phenomenological),
but rather is equivalent to a physical system with
an explicit energy function or Hamiltonian.

\section{Acknowledgments}

We gratefully acknowledge 
Dr.~Eli Hershkovitz for insightful discussions
and Dr.~Alexander Popov for a critical reading of the manuscript.
This work has been partially supported by a National Science Foundation 
Grant, No.~NSF 02-123320.  
The Center for Computational Science and Technology
is supported through a Shared University Research (SUR)
grant from IBM and Georgia Tech.
Additionally, RH is the Goizueta Foundation Junior Professor.


\bibliography{j,tst,miller2,hern,voth,cc,polymer,liquid,gas,md,bio,eli}



\end{document}